# A Scheme for Ultrasensitive Detection of Molecules by Using Vibrational Spectroscopy in Combination with Signal Processing

*Tay Yong Boon[1], Ian Tay Rongde[1], Loy Liang Yi[2‡], Aw Ke Fun[2‡], Ong Zhi Li[2‡], Sergei Manzhos[3,]*.*

[1] Raffles Institution, One Raffles Institution Lane, Singapore 575954.

[2] NUS High School of Mathematics and Science, 20 Clementi Avenue 1, Singapore 129957.

[3*] Department of Mechanical Engineering, National University of Singapore, Block EA #07-08, 9 Engineering Drive 1, Singapore 117576. Tel: +65 6516 4605; fax: +65 6779 1459. E-mail: mpemanzh@nus.edu.sg.

We show that combining vibrational spectroscopy with signal processing can result in a scheme for ultrasensitive detection of molecules. We consider the vibrational spectrum as a signal on the energy axis and apply a matched filter on that axis. On the example of a nerve agent molecule, we show that this allows detecting a molecule by its vibrational spectrum even when the recorded spectrum is completely buried in noise, when conventional spectroscopic detection is impossible. Detection is predicted to be possible with signal-to-noise ratios in recorded spectra as low as 0.1. We study the importance of spectral range used for detection as well as of the quality of the computed spectrum used to program the filter, specifically, the role of anharmonicity, of the exchange correlation functional, and of the basis set. The use of the full spectral range rather than of a narrow spectral window with key vibrations is shown to be advantageous, as well as accounting for anharmonicity.

**Introduction**

Detection of molecules is important for civil as well as defense applications. This includes detection of harmful molecules such as chemical warfare (CW) agents (or their precursors or reaction / decay products) or molecules which are indicative of presence of other controlled substances (explosives, fuels etc.). Such detection is desirable at the lowest possible concentrations even when such concentrations are not per se harmful. Extremely low concentrations can be critically harmful, for example, the LC50 dose (the lethal concentration required to kill 50% of the population) of $AsF_5$ gas is only 20 ppm; it is desirable be able to detect harmful molecules at much smaller concentrations than LC50. While multiple detection techniques exist, they all have their limitations, specifically, in sensitivity and selectivity. For example, detection of molecules by light absorption / photoluminescence properties has been proposed[1] but is very non-selective, as many molecules absorb or luminesce in similar bands. Some methods require a liquid sample, which may not be available; for example, when detecting molecules in the atmosphere or at the border of a territory where release may occur. As stated in a relatively recent review,[2] " [available] devices have several limitations, such as low specificity and inability to detect all CW agents. Definitive identification of an agent can be carried out onsite in a mobile analytical laboratory or in an off-site laboratory, and this will generally take many hours. Clinical symptoms and signs in exposed individuals may be the most useful indicators of the likely agent." The last sentence highlights how important detection is and that available methods are still deficient. One certainly does not want to be in a situation where "Clinical symptoms and signs in exposed individuals may be the most useful indicators of the likely agent".

IR (infrared) detection and other vibrational spectroscopies (for example those which are designed to detect adsorbed molecules such as surface-enhanced Raman etc.) can be very sensitive and selective.[3] Vibrational spectroscopy is a workhorse characterization technique used for species identification in many applications, from atmospheric chemistry to heterogeneous catalysis to Li ion batteries[4]. It can work in different environments, including ambient air. For example, one can record an IR spectrum by passing an IR laser beam through a relatively extended path in the atmosphere, e.g. by using surveillance aircraft. Vibrational spectroscopy is typically used to monitor a small number of characteristic frequencies of specific functional groups such as –COO, -OH[4-6] etc. However, similar vibrational frequencies can result from the same or other functional groups in very different molecules, especially if molecules interact among themselves or with a substrate.[3,4] This makes assignment unreliable, and multiple other pieces of information and intuition are necessary for a reliable assignment. Also, a good SNR (signal-to-noise ratio) (certainly >1) of recorded spectra is necessary. That is to say, selectivity and sensitivity are limited.

In the gas phase, when there is little interaction between molecules, one can match the measured spectrum with spectra in the library to detect a molecule. Only in the gas phase does one have the luxury of having a quality reference spectrum. In that case, one can achieve sub-ppm-level sensitivity;[7] it is achieved by using a large number of vibrational transitions (in a wide frequency range) and not only those due to fundamental transitions in key functional groups. The sensitivity drops to a fraction of a wt% in liquid, where the transition frequencies are perturbed/broadened by the environment and where the considered frequency range is typically smaller.[8]

Note that one may not have the ability to measure a quality reference spectrum because the chemical can be dangerous or restricted or because the spectrum cannot be measured well (with sufficient resolution, spectral range, and SNR) in the target environment. This is the case of the molecule considered here. In this case, a sufficiently accurate computed spectrum may be used as reference. Such spectra are computable by using an ab initio approach coupled with a method to solve the vibrational Schrödinger equation. The computed spectra can be perturbed/broadened to mimic environmental effects. For small molecules, various wavefunction based methods can be used, while for larger molecules, DFT (density functional theory) is often the only practical option.[9] We will use DFT in this article, which is sufficient for the purpose of this work. Vibrational spectra are easily computed in the harmonic approximation; accurate spectra, however, require treatment of anharmonicity and coupling, either perturbatively[10] or by solving the vibrational Schrödinger equation using an analytic potential energy surface representation[11,12] or its discreet samples.[13-17] The perturbative approach is implemented in major ab initio codes and can provide good accuracy in the absence of resonances, it is thus easily applicable for applications and is used here, while methods to properly solve the vibrational Schrödinger equation require more CPU and manpower resources.

In this article, we propose and computationally test the concept of ultrasensitive detection using the vibrational spectrum in a wide frequency range coupled with optimal filtration. The vibrational spectrum of a molecule is not limited to lowest-quanta transitions of a few key functional groups' models. Considered from the point of view of signal processing, the complete vibrational spectrum (absorption intensity as a function of frequency) even of a small molecule is a very complex signal. For example, while the vibrational spectrum of phosgene (a famous CW agent) is dominated by several strong peaks,[18] there are actually 250 transitions just up to 2,700 cm$^{-1}$;[15] many of those transitions are IR inactive, but many are and have small intensity. The complexity of the entire vibrational spectrum considered as a signal means that it is a unique molecular fingerprint. This signal complexity (and therefore a sharp autocorrelation function) also means that its recovery in the presence of noise could be efficient with signal processing techniques[19] such as matched filtering. This in turn means that selectivity and sensitivity could be drastically improved vs existing techniques. To use this idea, one must have accurate benchmarks (reference spectra) of vibrational spectra in a wide excitation energy range to program the matched filter. The reliance of the detector on multiple spectral lines in a wide frequency range means that anharmonicity should be considered when using computed spectra as reference.

Here, we apply this proposed scheme to detect the vibrational spectrum of the nerve agent A232 (the lay name "Novichok") in the presence of noise. A232, whose molecular structure is shown in Figure 1, is an organophosphorus compound which is deadly in tiny concentrations. It inhibits the enzyme acetylcholinesterase and cripples the nervous system.[20] While the molecular structure is known, characterization data are not easily available. We compute the IR spectrum of this molecule with different DFT based approaches, including different functionals, basis sets, and treatment of anharmonicity. We use the highest quality computed spectrum mixed with noise to model the measured spectrum and spectra computed with all considered levels of theory to program a matched filter in the frequency space. In this way, we study the effect on detection limit of the quality of the computed spectrum (effect of the functional, the basis, and of the harmonic vs anharmonic approximation) and of the width of the spectrum window used for detection. We find that the molecule can be detected even when the noise makes the spectrum visually completely unrecognizable. As expected, we find that lower levels of theory lead to higher detection limits (higher required SNR). We specifically find that anharmonic calculations are much preferred for use as reference spectra, highlighting the practical value of highly accurate computational anharmonic spectroscopy.

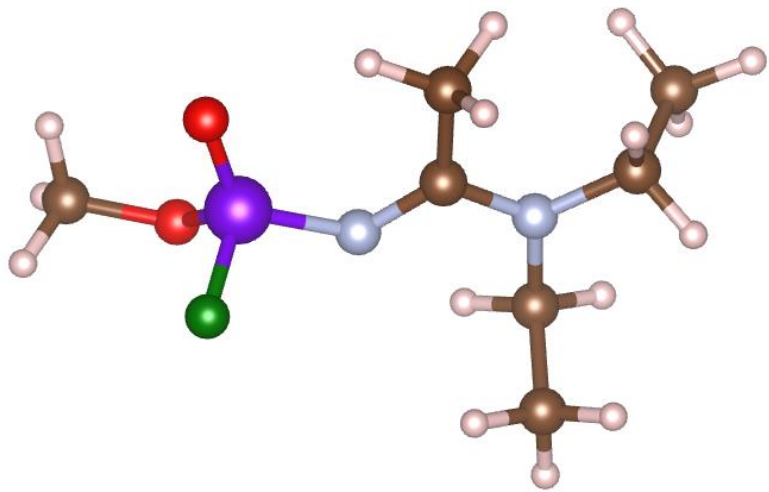

**Figure 1**. The molecular structure of the A232 nerve agent. Atom color scheme: C – brown, H – pink, N – blue, O – red, P –violet, F - green.

## Methods

DFT[21,22] calculations were performed in Gaussian 09[23] using PBE[24] and B3LYP[25] functionals. Different basis sets were used: 6-31g and 6-311g+(2d,2p). Basis sets of intermediate size were also tried but did not produce new knowledge. Different possible conformers of the A232 molecule were tried and the lowest energy conformer (shown in **Figure 1**) was used in subsequent calculations. Vibrational spectra were computed in the harmonic and the anharmonic approximations with all (four) combinations of functionals and basis sets. The anharmonic calculations were performed using the 2$^{nd}$ order perturbation theory. The spectra were Gaussian-broadened by 1 cm$^{-1}$ (Gaussian width) for further processing.

The matched filter was coded in Octave[26]. A matched filter[27] is a linear filter that maximizes the SNR in the presence of a given type of noise. For a discrete signal $x'_k = x'(k)$, where $k$ indices discretization points, the output $y_n = y(n)$ is computed as

$$y_n = \sum_{k=-\infty}^{\infty} h(n-k) x'_k$$

(1)

where $h$ is the filter. The matched filter correlates the received signal (a vector *x'* indexed by $k$) with a filter *h* (another vector) that is parallel with the signal, maximizing the inner product. This is achieved when $h = \alpha R_\xi^{-1} x$, where $R_\xi$ is the covariance of the noise and $\alpha$ a normalization constant. The vector $x$ is the useful (expected) signal component of the input $x'$ which is deteriorated by the noise $\xi$: $x' = x + \xi$: For white noise assumed here, we can put $h = x$ and matched filtering becomes equivalent to correlating the received signal with the expected signal with the output $y$ proportional to the signal's autocorrelation function. The technique is therefore the more powerful the more complex the shape of $x$ is i.e. the sharper its autocorrelation function, ideally approaching the delta function for very complex-shaped signals. The quality of the detection deteriorates if there is mismatch between $h$ and $x$. A popular example is relatively easy radar detection of stealth aircraft using complex-shaped, broadband signals. In signal processing applications, $k$ usually indexes time intervals.

In this work, we consider the vibrational spectrum as such a complex signal with a sharp autocorrelation function when considered along the frequency axis. We therefore apply the matched filter along the frequency axis. The highest quality computed spectrum (that with B3LYP functional and 6-311g+(2d,2p) basis set and with the anharmonic corrections) is used to emulate the signal $x$; it plays the role of a benchmark. It is mixed with white noise and is detected using a spectrum computed with different levels of theory as $h$. We also test performing the detection in the entire relevant spectral range (set here from 0 to 4,000 cm$^{-1}$) vs. a frequency window (from 680 to 2000 cm$^{-1}$).

The frequency axis was discretized with a step of 0.061 cm$^{-1}$ using $2^{16}$ points ($k$ values) up to 4,000 cm$^{-1}$. Under spectral broadening of 1 cm$^{-1}$ used here, this provided converged results (no changes were detected with a finer discretization). The white noise was subjected to the same broadening. Unsigned noise was added to the spectrum for computational simplicity. The peak in the autocorrelation function was detected by looking for any elements of $y$ which exceed $N \times \sigma$ where $\sigma$ is the standard deviation of $y$ and $N$ is a chosen parameter. The detection limit is defined as a noise level with which the rate of false negatives reaches 50%. False positives in this case are due to a finite probability of satisfying the criterion $y_k > N \times \sigma$ with the noise in the absence of the signal $x$. To reduce the rate of false positives, the peak search is performed in the window of width 300 and 80 cm$^{-1}$ (when using the full spectrum range and the window 680 - 2,000 cm$^{-1}$, respectively) around the mid-point of $y$. When $h = x$ (i.e. when the same level of theory is used for the signal and for the filter), the peak is exactly at the mid-point. When different levels of theory are used, the peak may be off-center; the peak detection window was chosen for the worst mismatch among all cases. The rate of false positives was computed by using noise-only inputs. With each instance of noise (for both signal-containing and noise-only inputs), 50 numeric experiments were performed. The corresponding SNR is defined as the ratio of the root mean square of the signal $x$ (the spectrum) and the standard deviation of the noise, both under the same broadening of 1 cm$^{-1}$.

**Results**

Molecular geometry computed at the B3LYP/6-311g+(2d,2p) level of theory as well as vibrational spectra computed with all combinations of functionals and basis sets are available upon request. **Figure 2** shows computed IR spectra for several combinations of basis-functional, harmonic and anharmonic. **Figure 3** shows the (benchmark) spectrum computed at the highest level of theory (B3LYP/anharmonic/6-311+g(2d,2p)) with different levels of noise. (The units of intensity, which are 10$^{-40}$ esu$^2$ cm$^2$, are dropped in the subsequent discussion as they are unimportant and make no difference to the conclusions as long as all calculations are done at the same scale and with the same broadening, which is the case). At noise levels on the order of 500 the spectrum is completely unusable for assignment. This case corresponds to a value of SNR as defined above of about 0.25. Examples of correlation functions (filter output $y$) of the benchmark spectrum with spectra computed at different levels of theory are shown in **Figure 4**. The top pair of plots show the autocorrelation function which possesses a sharp peak in the middle, which is due to the complex shape of $x$ and which can permit detection of the molecule in the presence of high levels of noise. The following three pairs of panels show the effect of errors in the computed spectrum (which programs the filter $h$) on the quality of the peak in $y$ due to the use of the harmonic approximation, of a less accurate exchange-correlation functional, and of a small basis set, respectively. The final pair of plots (at the PBE/harmonic/6-31g level) show the combined effect of these sources of error. Note that the application of Eq. (1) results in the range of the abscissa values for $y$ which is double that of the original spectrum, i.e. 8,000 cm$^{-1}$, with the autocorrelation peak centered at 4,000 cm$^{-1}$.

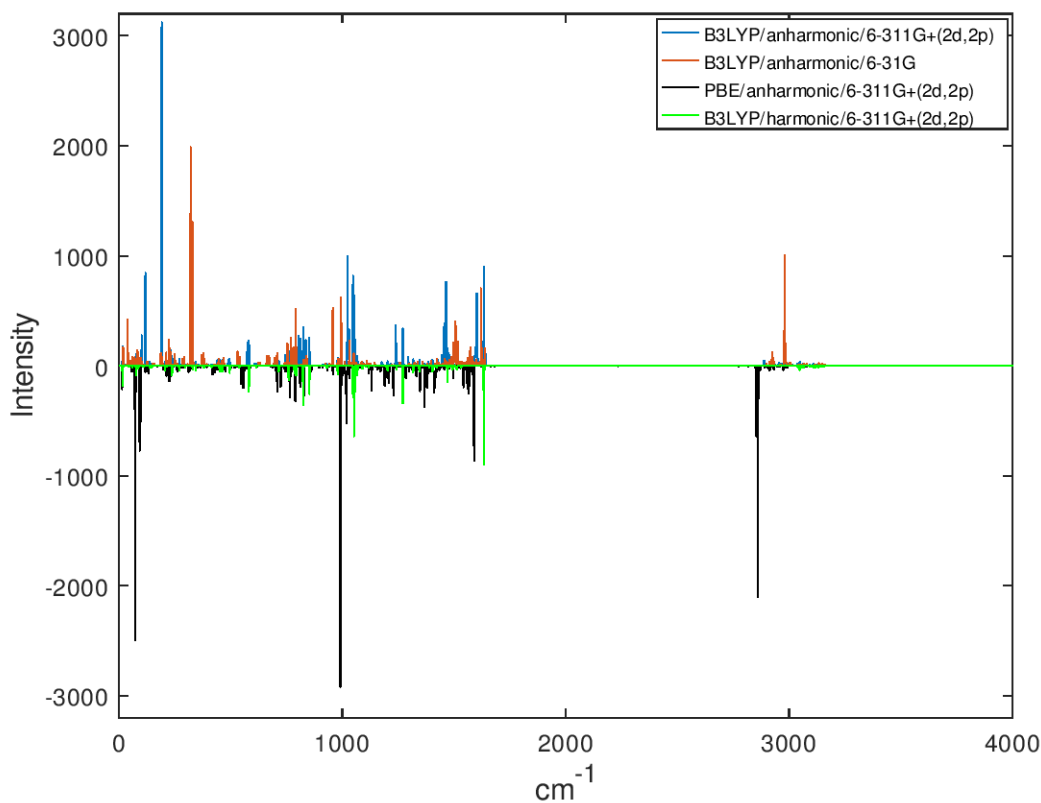

**Figure 2**. IR spectra of the A232 molecule computed with different levels of theory. Spectra are Gaussian broadened by 1 cm$^{-1}$ (Gaussian width). Some spectra are plotted in the negative for better visibility. The intensities are in $10^{-40}$ esu$^2$ cm$^2$.

As expected, lower levels of theory deteriorate the quality of the output of the filter (height and position of the main peak used for detection). The effects of various components of the error are different. For example, use of the harmonic approximation (other approximations being equal) reduces the peak height by about an order of magnitude but does not shift significantly its position. The use of the PBE functional (other parameters being equal) reduces the peak height "only" by a factor of 2 but shifts its position by more than 100 cm$^{-1}$ with respect to the position to the autocorrelation function's peak. Qualitatively, a similar effect is observed when using a smaller basis (other parameters being equal). All lower-level approximations lead to the appearance of strong off-center peaks. The combined effect of all these sources of error (bottom pair of plots) is a peak height reduced by about a factor of 20 and shifted by more than 100 cm$^{-1}$, with multiple satellite peaks, some exceeding in height the central peak. Overall the results shown in **Figure 4** suggest that anharmonicity treatment is important as well as proper choices of functional and basis set, and that one must allow for a "window" on the order of ±200 cm$^{-1}$ for peak detection.

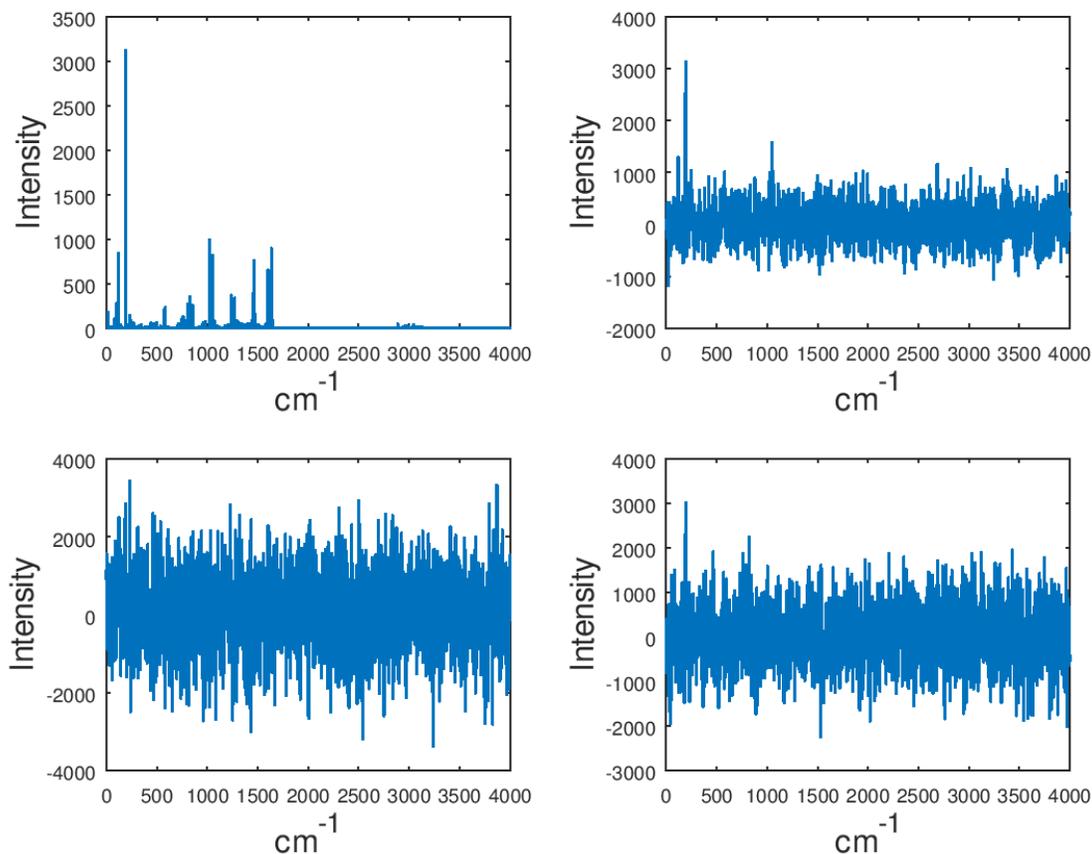

**Figure 3**. The IR spectrum of A232 computed at the B3LYP/6-311+g(2d,2p) level and summed with white noise of amplitudes (clockwise from top left) 0, 500, 1000, and 1500.

We also explored detection in a selected spectral window of 680-2000 cm$^{-1}$. In this case, components of *x'* and of *h* outside this range were set to zero. The resulting correlation functions (filter output *y*) of the benchmark spectrum with spectra computed at different levels of theory are shown in **Figure 5**. The use of a spectral window reduced by about a factor of 2 the peak height in the autocorrelation function, but it also reduces the degree of relative deterioration of the main peak quality with approximate computational schemes. For example, the use of the harmonic approximation (other parameters being equal) reduced the peak height by about a factor of 2½. The use of the PBE functional (other parameters being equal) caused the peak's shift but hardly reduces its height.

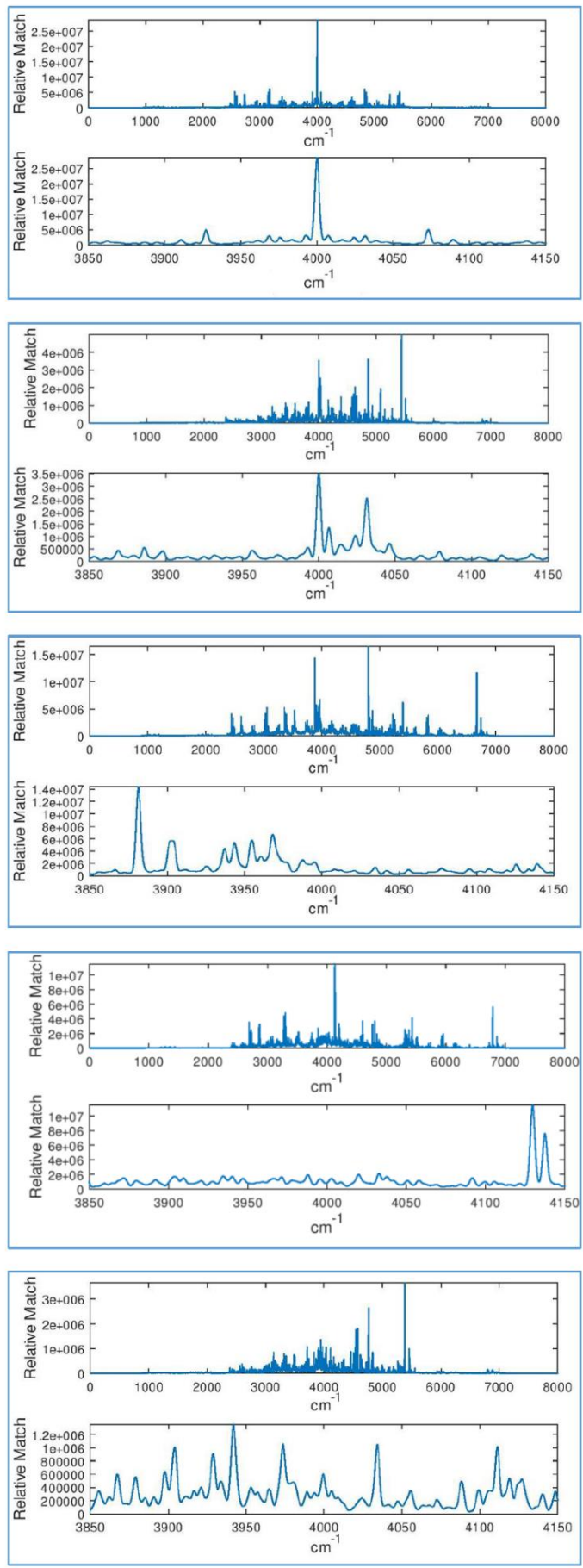

**Figure 4**. The correlation function of the reference spectrum (anharmonic spectrum computed with B3LYP/6-311+g(2d,2p)) with spectra computed at different levels of theory, using the full spectral range 0 – 4,000 cm$^{-1}$. From top to bottom: B3LYP/anharmonic/6-311+g(2d,2p) (i.e. autocorrelation), B3LYP/harm/6-311+g(2d,2p), PBE/anharmonic/6-311+g(2d,2p), B3LYP/anharmonic/6-31g, and PBE/harmonic/6-31g. The top panel in each pair of plots is for the entire signal y and the bottom panel is zoomed around the center, where a peak is expected.

The use of a small basis (other parameters being equal) however reduces the peak height by about a factor of 4 and has a worse effect than in the case of a full spectral range. The combined effect of all these sources of error is a peak weakened by an order of magnitude with multiple and strong satellite peaks. It may appear therefore based on **Figure 5** that using a limited spectral range may be beneficial if the filter is programmed with a computed spectrum using a sufficiently complete basis set.

We computed the detection limits and corresponding SNR ratios when detecting the benchmark spectrum deteriorated with different levels of noise with a filter programmed with a spectrum computed at different levels of theory, using the full spectrum up to 4,000 cm$^{-1}$ or a detection window of 680 – 2000 cm$^{-1}$. The detection thresholds were determined by comparing the rate of positives with the rate of false positives (on noise-only inputs) and correspond to 50% probability detection. Different $N$ were tried in the peak detection criterion $y_k > N \times \sigma$. This criterion was applied in the window of ±150 (when using full spectral range) or ±40 (when using a window 680 - 2,000 cm$^{-1}$) cm$^{-1}$, based on the results of **Figure 4**, **Figure 5**. Examples of these calculations when programming *h* with the spectrum computed at the B3LYP/anharmonic/6-311+g(2d,2p) level are shown in **Figure 6** for the case of full spectrum detection and in **Figure 7** for detection in the spectral window. The results for all cases are shown in **Table 1**. The results show that with the criterion $y > 6\sigma$, one can achieve a sufficiently low rate of false positives of <<0.1. When the filter response function is programmed with a high quality computed spectrum, the detection threshold SNR can then be as low as about 0.1, corresponding the "noisiest" panel of **Figure 3** which appears to consist entirely of noise. This is achieved by using the entire spectral range considered here (0 – 4,000 cm$^{-1}$). Detection in a narrower window, in spite of some advantages listed above, results in higher required SNR, albeit reliable detection should still be possible with SNR<0.5. This highlights the fact that it is really in a wide spectral range that the molecular vibrational spectrum has unique, fingerprint-like quality.

**Conclusions**

We explored computationally a scheme for detection of molecules that combines vibrational spectroscopy with optimal filtration. We considered the vibrational spectrum of a molecule in a wide spectral range as its unique fingerprint. The spectral shape presents a complex signal with a sharp autocorrelation function, which makes it suitable for detection using matched filtering. We showed that this scheme can be applied for detection of molecules at low concentrations where the signal-to-noise ratio of recorded spectra is very low. Specifically, we considered detection of a nerve agent A232 which is lethally harmful at trace concentrations and showed that the proposed scheme can detect the presence of the molecule when the SNR is as low as 0.1 i.e. when the recorded spectrum is completely buried in noise. A232 is also an example of a molecule where experimental reference spectra needed to code the matched filter are not accessible; this is true for most dangerous substances. The matched filter can be programmed by a computed spectrum. We studied the sensitivity of the scheme when different levels of theory are used to compute the spectrum for the matched filter such as different exchange-correlation functionals, basis sets, and treatment of anharmonicity. The sensitivity of the detector is, as expected, improved with higher levels of theory. Specifically, anharmonic calculations are shown to be much preferred; this highlights the practical value of computational anharmonic spectroscopy and suggests that methods properly solving the vibrational Schrödinger equation (as opposed to perturbative treatment used here) on the extended potential energy surface (either analytic or discretely sampled with ab initio calculations) should be explored for this application. In general, this work highlights the value of highly accurate computational vibrational spectroscopy. We also observed the advantage of considering a wide spectral range: detection in a narrower spectral window, in spite of some advantages, results in higher required SNR, albeit reliable detection should still be possible with SNR<0.5. This highlights the fact that it is really in a wide spectral range that the molecular vibrational spectrum has unique, fingerprint-like quality. In this work, we have ignored rotational contributions to the spectra. For relatively large molecules like the one considered here, it may not be practical to treat rotational lines individually, and appropriate broadening could be used instead. For sufficiently small molecules where rotational resolution is possible, the ro-vibrational spectrum can offer increased complexity and more delta function-like autocorrelation function which could lead to even more sensitive schemes.

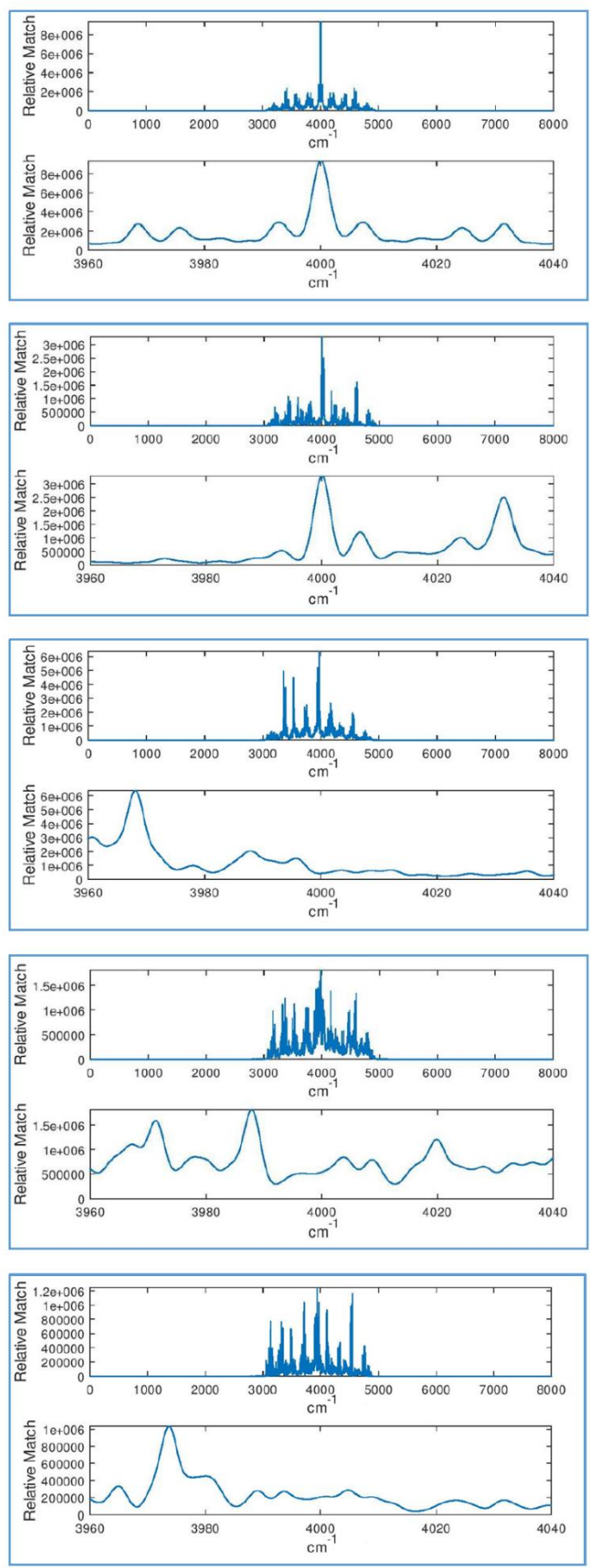

**Figure 5**. The correlation function of the reference spectrum (anharmonic spectrum computed with B3LYP/6-311+g(2d,2p)) with spectra computed at different levels of theory within a detection window of 680 – 2000 cm$^{-1}$. From top to bottom: B3LYP/anharm/6-311+g(2d,2p) (i.e. autocorrelation), B3LYP/harm/6-311+g(2d,2p), PBE/anharm/6-311+g(2d,2p), B3LYP/anharm/6-31g, and PBE/harm/6-31g. The top panel in each pair of plots is for the entire signal y and the bottom panel is zoomed around the center, where a peak is expected.

**Table 1.** Detection levels (noise levels and corresponding SNR values in parentheses) when using spectra computed at different levels of theory to program the filter. The reference spectrum computed at the highest level of theory is used to model the signal deteriorated by noise. Lower level approximations vs. the reference spectrum are highlighted in bold. The results are shown for most primising peak detection criteria ($N\times\sigma$).

| | Detection window | | Full spectrum | |
|---|---|---|---|---|
| | $N\times\sigma$ detection criteria | | | |
| | $4\sigma$ | $5\sigma$ | $5\sigma$ | $6\sigma$ |
| Method | Detection threshold: NoiseLevel (SNR) | | | |
| B3LYP \ anharmonic \ 6-311G+(2d,2p) | 819 (0.17) | 608 (0.22) | 1600 (0.08) | 1317 (0.10) |
| B3LYP \ **harmonic** \ 6-311G+(2d,2p) | 533 (0.25) | 388 (0.34) | 684 (0.20) | 465 (0.29) |
| B3LYP \ anharmonic \ **6-31G** | 250 (0.56) | 143 (0.97) | 869 (0.16) | 672 (0.20) |
| **PBE** \ anharmonic \ 6-311G+(2d,2p) | 373 (0.37) | 304 (0.46) | 700 (0.20) | 522 (0.26) |
| **PBE** \ **harmonic** \ **6-31G** | 190 (0.72) | 156 (0.85) | 357 (0.38) | 221 (0.62) |

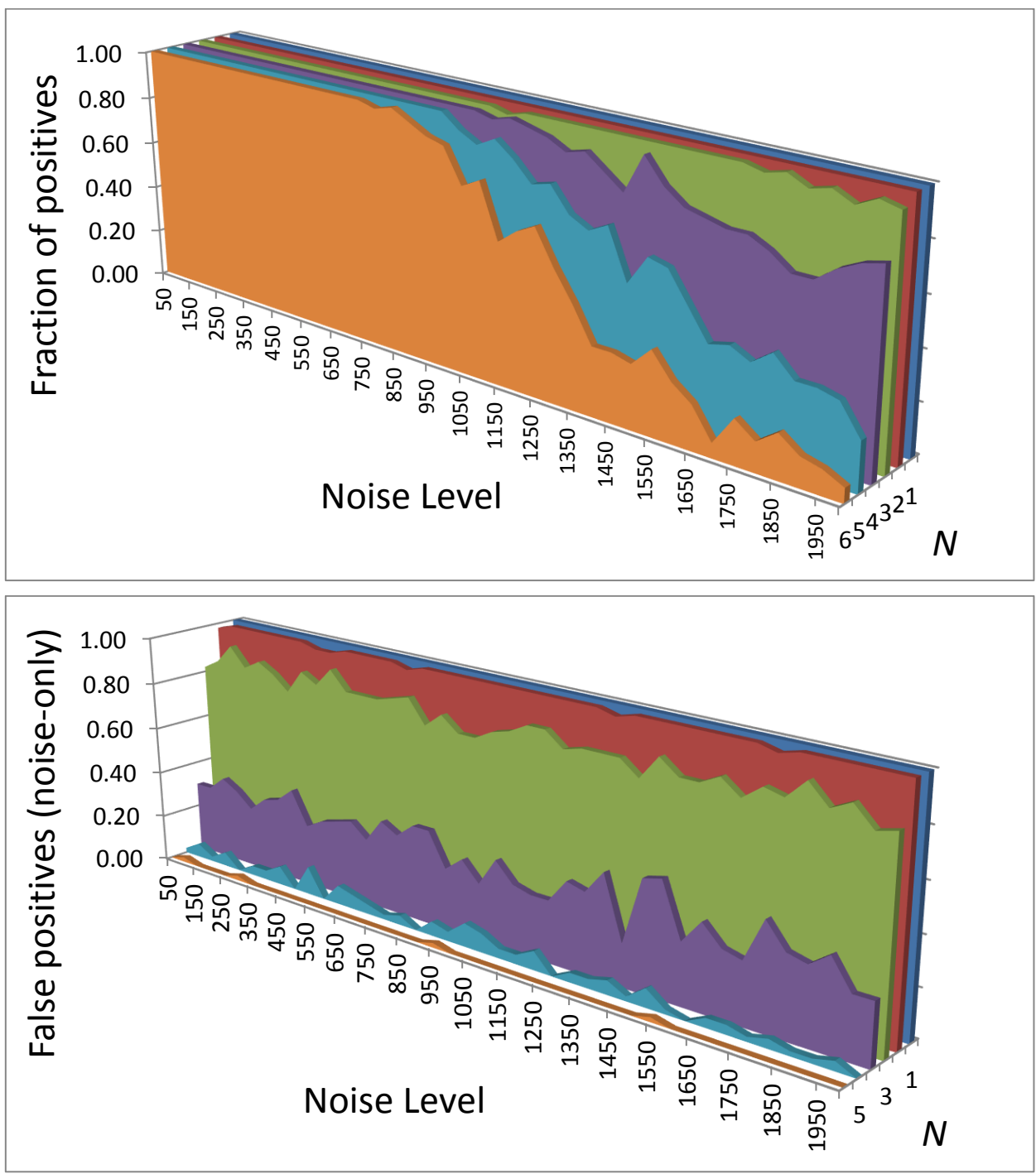

**Figure 6**. The fraction of positive detection outcomes (detection criterion $y > N\sigma$ satisfied) for different levels of noise and $N$, for detection in the full spectral range of 0 – 4,000 cm$^{-1}$. The top panel is for the inputs consisting of the molecular spectrum and noise and the bottom panel for noise-only inputs. The filter is programmed with a spectrum computed at the B3LYP/anharm/6-311+g(2d,2p) level.

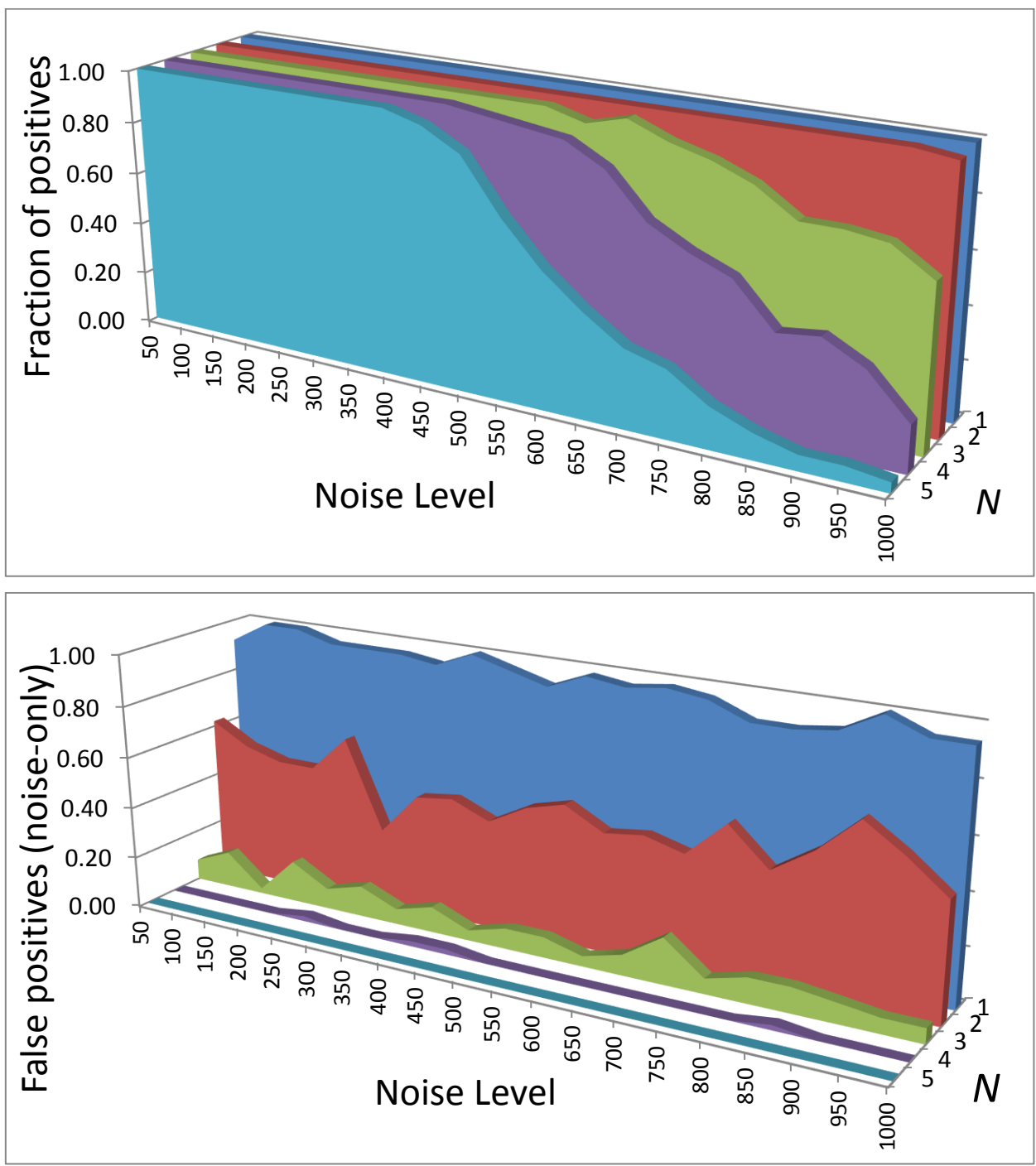

**Figure 7**. The fraction of positive detection outcomes (detection criterion $y > N\sigma$ satisfied) for different levels of noise and $N$, for detection in the full spectral window $680 - 2{,}000$ cm$^{-1}$. The top panel is for the inputs consisting of the molecular spectrum and noise and the bottom panel for noise-only inputs. The filter is programmed with a spectrum computed at the B3LYP/anharm/6-311+g(2d,2p) level.

**Author Contributions**

The manuscript was written through contributions of all authors. All authors have given approval to the final version of the manuscript. ‡These authors contributed equally. S.M.: conceptualization, project management, and text. T.Y.B., I.T.R., L.L.Y., A.K.F., O.Z.L.: calculations and data processing.

**Acknowledgement**

The authors also thank the Science Enrichment Program of the National University of Singapore.